\documentclass[aps,twocolumn,groupedaddress]{revtex4}
\usepackage{epsfig}
\usepackage{graphicx}
\usepackage[T1]{fontenc}
\usepackage{ae}
\usepackage{color}
\usepackage[latin1]{inputenc}

\begin{document}



\title[Dynamics of magnetic dipoles]
{Dynamics and thermodynamics of a pair of interacting magnetic
dipoles}

\author{Heinz-J\"urgen Schmidt$^1$
\footnote[3]{Correspondence should be addressed to hschmidt@uos.de}
, Christian Schr\"oder$^2$, Eva H\"agele$^2$ and Marshall Luban$^3$ }

\affiliation{$^1$Department of Physics, University of Osnabr\"uck,
 D - 49069 Osnabr\"uck, Germany\\
$^2$Department of Engineering
Sciences and Mathematics, University of Applied Sciences,
D - 33692 Bielefeld,
Germany \\
$^3$Department of Physics and Astronomy, Iowa State University, Ames, IA 50011,
USA }


\begin{abstract}
We consider the dynamics and thermodynamics of a pair of
magnetic dipoles interacting via their magnetic fields. We consider only the ``spin" degrees
of freedom; the dipoles are fixed in space. With this restriction it is possible to
provide the general solution of the equations of motion in analytical form.
Thermodynamic quantities, such as the specific heat and the zero field susceptibility are calculated
by combining low temperature asymptotic series and a complete high temperature expansion.
The thermal expectation value of the autocorrelation function is determined for the
low temperature regime including terms linear in  $T$.
Furthermore, we compare our analytical results with numerical calculations based on Monte Carlo simulations.
\end{abstract}

\maketitle

\section{Introduction}\label{sec:I}
Systems in which magnetic nanostructures solely interact via
electromagnetic forces have recently drawn much attention
experimentally as well as theoretically \cite{EW13} - \cite{JO98}.
Whereas in traditional magnetic systems electromagnetic forces
usually just add to a complex exchange interaction scenario, they
play a major role in arrays of interacting magnetic nanoparticles
and lithographically produced nanostructures. In such systems
geometrical frustration and disorder lead to interesting and exotic
low temperature effects, e.~g.~artificial spin ice  \cite{W06},
\cite{C08}, and superspin glass behavior \cite{H11}. Moreover, these
systems are promising candidates for future applications beyond
magnetic data-storage, e.~g.~, as low-power logical devices
\cite{I06}, \cite{E14}. Theoretically, these systems can often be described as
interacting point dipoles. This is justified if the considered
nanostructures form single domain magnets and are spatially well
separated from each other so that exchange interactions do not play
an important role. In this paper, we show that the dynamical and
thermodynamical properties of the basic building block of such
systems, a pair of interacting point dipoles, can rigorously be
treated analytically by combining low temperature asymptotic series
and a complete high temperature expansion. A considerable part of these
calculations has been performed with the aid
of the computer algebra system MATHEMATICA 9.0.\\

From a mathematical point of view, the system of two interacting
magnetic dipoles is equivalent to a classical spin system
with $N=2$ and the particular XXZ Hamiltonian (\ref{DS2a}).
Hence our results can be applied to these systems as well.\\

The paper is organized as follows. For the reader's convenience
we recapitulate in section \ref{sec:DE} the derivation of the
equation of motion (eqm) of two interacting dipoles and identify the
underlying assumptions. The solution of the eqm in terms of elliptic
integrals and the Weierstrass elliptic function in section
\ref{sec:DS} is based on the existence of two conserved quantities.
The limiting case of solutions close to the ground state can be
described by harmonic oscillations with three frequencies, see
section \ref{sec:DL}. In the next sections we discuss the
thermodynamics of the dipole pair.
After explaining our methods we calculate the partition
function (section \ref{sec:Z}), the specific heat (section
\ref{sec:CH}) and the zero field susceptibility (section
\ref{sec:Chi}) by combining low- and high-temperature expansions.
The latter two physical properties are also determined by Monte Carlo
simulations and shown to closely coincide with the theoretical
results. Since the problem is anisotropic we have to distinguish
between different susceptibilities w.~r.~t.~the ``easy axis", the
axis joining the two dipoles, and the ``hard axis", any axis
perpendicular to the easy axis. For the easy axis susceptibility
there occur complications for the standard Monte Carlo simulations that have
been overcome by using the so-called Exchange Monte Carlo method,
see \cite{H96}. Similarly the autocorrelation function is calculated
in the low temperature limit and compared with simulation results at
low temperatures, see section \ref{sec:AC}. We find that one of the
three frequencies mentioned above is suppressed by thermodynamical
averaging. Appendix A contains a short introduction into the theory
of elliptic integrals and elliptic functions for those readers who
are not acquainted with this subject. The Appendices B -- D contain
details of the theoretical derivations presented in the main part of
the paper. We close with a summary and outlook.

\section{Dynamics}
\label{sec:DY}
\subsection{Derivation of the equation of motion}\label{sec:DE}
We consider two identical magnetic dipoles, labeled by an index  $i=1,2$, that are fixed in space
and separated by a distance $a$.
We denote the magnetic moment vector of dipole $i$ by ${\mathbf m}_i$
and assume that it is associated with an angular momentum ${\mathbf L}_i$
according to the standard formula
\begin{equation}\label{DE1}
{\mathbf m}_i= \gamma \;{\mathbf L}_i\;, \quad i=1,2
\;,
\end{equation}
where $\gamma$ is the gyromagnetic ratio
\begin{equation}\label{DE2}
\gamma = - \frac{e_0}{2 m_e} g
\;,
\end{equation}
independent of $i$. $\gamma$ is assumed to be negative due to the negative charge $-e_0$ of the electron
($m_e$ denoting its mass)
and the gyromagnetic factor $g$ is considered as a physical property of the dipoles. We expect that
$g$ varies between $g=1$ for the contribution due to pure orbital motion of the electrons and $g=2$ for
the spin contribution to the magnetism of the dipoles. Furthermore we will assume that the torque exerted on a dipole
by a magnetic field ${\mathbf B}$ is equal to
\begin{equation}\label{DE3}
{\mathbf N}= {\mathbf m} \times {\mathbf B}
\;.
\end{equation}
This textbook equation is usually derived for systems of moving charges and constant magnetic fields.
Hence the validity of (\ref{DE3}) for the problem under consideration is not trivial but an additional assumption.
In our case there are two magnetic fields, ${\mathbf B}_1$ and ${\mathbf B}_2$, where
${\mathbf B}_2$ denotes the instantaneous value of the magnetic field at ${\mathbf m}_1$ due to
dipole $2$, and an analogous definition applies for ${\mathbf B}_1$ due to dipole $1$. Thus, for example,
\begin{equation}\label{DE3a}
{\mathbf B}_2= \frac{ \mu_0}{4\pi a^3}\left( 3 {\mathbf m}_2 \cdot {\mathbf e}\;{\mathbf e}-{\mathbf m}_2 \right)
\;,
\end{equation}
where ${\mathbf e}$ is a unit vector parallel to the constant position vector from dipole $1$ to dipole $2$.
Hence we obtain
\begin{eqnarray}\nonumber
\frac{d}{dt}{\mathbf m}_1&=& \gamma\;\frac{d}{dt}{\mathbf L}_1 =\gamma{\mathbf m}_1 \times {\mathbf B}_2\\
\label{DE4}
&=&\frac{\gamma \mu_0}{4\pi a^3}{\mathbf m}_1\times\left( 3{\mathbf m}_2 \cdot {\mathbf e}\; {\mathbf e}-{\mathbf m}_2\right)
\;.
\end{eqnarray}

Introducing the unit vectors ${\mathbf s}_i=\frac{1}{M}{\mathbf m}_i,\;i=1,2$ where $M=|{\mathbf m}_1|=|{\mathbf m}_2|$
is constant, and utilizing (\ref{DE2}) we rewrite (\ref{DE4}) as
\begin{eqnarray}\label{DE5a}
\frac{d}{dt}{\mathbf s}_1&=&
-\frac{\mu_0 e_0\, g\, M}{8\,\pi\, m_e \,a^3}\;
{\mathbf s}_1\times\left( 3{\mathbf s}_2 \cdot {\mathbf e}\; {\mathbf e}-{\mathbf s}_2\right)\\
\label{DE5b}
&=& -\omega_0 \;{\mathbf s}_1\times\left( 3{\mathbf s}_2 \cdot {\mathbf e}\; {\mathbf e}-{\mathbf s}_2\right)
\;.
\end{eqnarray}
Here we have introduced the constant $\omega_0$, with dimension 1/time, defined by
\begin{equation}\label{DE5c}
\omega_0\equiv \frac{\mu_0 e_0\, g\, M}{8\,\pi\, m_e \,a^3}
\;.
\end{equation}
Using $\omega_0 t$ as a dimensionless time variable, again denoted by $t$, and considering the analogous
equation of motion (eqm) for the
second dipole, we eventually obtain the following system of coupled first order differential equations:
\begin{eqnarray}\label{DE6a}
\frac{d}{dt}{\mathbf s}_1&=&
-
{\mathbf s}_1\times\left( 3{\mathbf s}_2 \cdot {\mathbf e}\; {\mathbf e}-{\mathbf s}_2\right)
\;,\\
\label{DE6b}
\frac{d}{dt}{\mathbf s}_2&=&
-
{\mathbf s}_2\times\left( 3{\mathbf s}_1 \cdot {\mathbf e}\; {\mathbf e}-{\mathbf s}_1\right)
\;.
\end{eqnarray}

In view of possible applications mentioned in the Introduction we stress that the
derivation of the eqm (\ref{DE6a}), (\ref{DE6b}) is based on the following two idealized assumptions:
\begin{itemize}
\item The two dipoles can be assumed as point-like objects, and
\item the constant $\omega_0$ is small enough such that the quasi-static approximation of
the complete set of Maxwell's equations is valid.
\end{itemize}

\subsection{Solution of the equation of motion}\label{sec:DS}

To facilitate solving the eqm (\ref{DE6a}), (\ref{DE6b}) we first note that these
equations give rise to two conserved physical quantities, to be denoted by $Q_1$ and $Q_2$:
\begin{equation}\label{DS1}
Q_1={\mathbf S}\cdot {\mathbf e}
\;,
\end{equation}
where ${\mathbf S}\equiv {\mathbf s}_1+{\mathbf s}_2$,
and $Q_2$ is the dimensionless energy
\begin{equation}\label{DS2}
Q_2=H
\;,
\end{equation}
that can be written in any one of the following four forms
\begin{eqnarray}\label{DS2a}
H&=&- {\mathbf s}_1 \cdot \left( 3{\mathbf s}_2 \cdot {\mathbf e}\; {\mathbf e}-{\mathbf s}_2\right)\\
\label{DS2b}
&=&- 3{\mathbf s}_1 \cdot {\mathbf e}\;{\mathbf s}_2 \cdot {\mathbf e}+{\mathbf s}_1 \cdot {\mathbf s}_2\\
\label{DS2c}
&=& -\frac{1}{E_0}{\mathbf m}_1\cdot {\mathbf B}_2= -\frac{1}{E_0}{\mathbf m}_2\cdot {\mathbf B}_1
\;.
\end{eqnarray}
Here we have introduced the unit of energy
\begin{equation}\label{DS2d}
E_0=\frac{\mu_0 \,M^2}{4\pi\, a^3}
\;.
\end{equation}
The quantity $Q_1$ is proportional to the component of the total magnetic moment in the
direction of ${\mathbf e}$ and obviously conserved due to the azimuthal symmetry of the problem in the spirit of Noether's theorem.
Moreover, from (\ref{DS2c}) it is clear that $Q_2$ is proportional to the total energy of the magnetic field
originating in the pair of dipoles. Its conservation reflects the time-translational symmetry of the problem. \\

It can be shown that (\ref{DS2a}) is the Hamiltonian for the
system (\ref{DE6a}),(\ref{DE6b}) as well in the sense of classical
mechanics. More precisely, we consider (\ref{DE6a}),(\ref{DE6b}) as
an eqm on the $4$-dimensional phase space
${\mathcal S}^2 \times {\mathcal S}^2$
with canonical coordinates $(p_i,\,q_i)=(\phi_i,\,z_i),\,i=1,2$ defined
by
\begin{equation}\label{DS3}
{\mathbf s}_i=
\left(
\begin{array}{l}
\sqrt{1-z_i^2}\cos \phi_i\\
\sqrt{1-z_i^2}\sin \phi_i\\
z_i
\end{array}
\right)
\;,
\end{equation}
where the $z$-axis has been chosen in the direction of ${\mathbf e}$,
and rewrite (\ref{DE6a}),(\ref{DE6b}) in the following form:
\begin{eqnarray}\label{DS4a}
\dot{\phi_1}&=& \frac{z_1
\sqrt{1-z_2^2} }   {\sqrt{1-z_1^2}}\cos (\phi_1-\phi_2)+2 z_2\;,\\
\label{DS4b}
\dot{\phi_2}&=& \frac{z_2
\sqrt{1-z_1^2} }   {\sqrt{1-z_2^2}}\cos (\phi_2-\phi_1)+2 z_1\;,\\
\label{DS4c}
\dot{z_1}&=& \sin (\phi_2-\phi_1)\sqrt{1-z_1^2}\sqrt{1-z_2^2}\;,\\
\label{DS4d}
\dot{z_2}&=& \sin (\phi_1-\phi_2)\sqrt{1-z_2^2}\sqrt{1-z_1^2}
\;,
\end{eqnarray}
where the dot denotes the derivative w.~r.~t.~time $t$.\\

\begin{figure}
\begin{center}
\includegraphics[clip=on,width=110mm,angle=0]{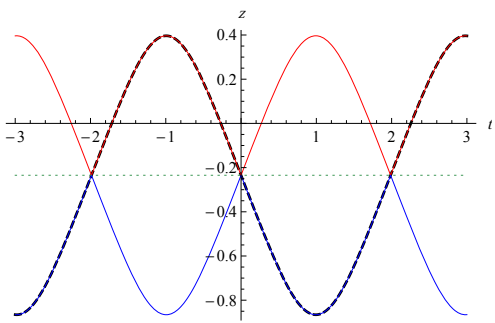}
\end{center}
\caption{Plot of a periodic solution $z_{1,2}(t)$ of (\ref{DS4a})--(\ref{DS4d}) with randomly chosen initial conditions.
The solution has been redrawn such that $z_{1,2}(t)$ assumes its mean value (dotted line) at $t=0$.
The black dashed curve represents the numerical solution $z_1(t)$
and the blue/red curves, partially hidden, the analytical solutions (\ref{DS7g}) according to the sign $\pm$.
\label{fig_z12}
}
\end{figure}

\begin{figure}
\begin{center}
\includegraphics[clip=on,width=110mm,angle=0]{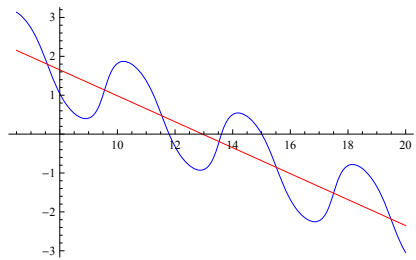}
\end{center}
\caption{Plot of a numerical solution $\phi_1(t)$ of (\ref{DS4a})--(\ref{DS4d}) with the same initial conditions as in figure \ref{fig_z12}.
One notes a constant drift superimposed by a periodic oscillation.
\label{fig_phi1}
}
\end{figure}

As a function of the canonical coordinates $H$ assumes the form
\begin{equation}\label{DS5}
H= \sqrt{1-z_1^2}\sqrt{1-z_2^2}\cos(\phi_1-\phi_2)-2 z_1 z_2
\;.
\end{equation}
Then it follows that
\begin{eqnarray}\label{DS6a}
\dot{\phi_i}&=& -\frac{\partial H}{\partial z_i}\;, i=1,2\;,\\
\label{DS6b}
\dot{z_i}&=& \frac{\partial H}{\partial \phi_i}\;, i=1,2
\;.
\end{eqnarray}

It can be shown that the solution $(\phi_1(t),z_1(t),\phi_2(t),z_2(t))$ of (\ref{DS4a})-(\ref{DS4d}) moves
on a $2$-dimensional torus defined by the equations ${\mathbf S}\cdot{\mathbf e}=s_3$ and $H=e$.
Since the number of conserved quantities is half the phase space dimension
the system (\ref{DS6a}), (\ref{DS6b}) is {\em completely integrable}
in the sense of the Arnol'd--Liouville theorem \cite{A78} and its solution can be
implicitly expressed in terms of integrals. In our case, these integrals are of elliptic
kind and hence the solution can be explicitly given by means of Weierstrass elliptic
functions ${\mathcal P}$ and elliptic integrals, see \cite{AS72} Ch. $17$ and $18$.
We will give some more details
of these calculations in Appendix \ref{sec:B} as well as a short introduction to the theory of
elliptic integrals and functions in Appendix \ref{sec:A}.
Here we will immediately formulate the final result for $z_{1,2}(t)$ after defining the quantities
\begin{eqnarray}\nonumber
g_2&\equiv&\frac{4}{27} \left(112 e^2+16 e \left(3 s_3^2-16\right)+9 s_3^4-168 s_3^2+208\right),\\
&&\label{DS7a}\\
\nonumber
g_3&\equiv&\frac{8}{729} \left(8 e+3 s_3^2-4\right)\\
\label{DS7b}
&& \left(80 e^2-48 e s_3^2-512 e-9 s_3^4-408
s_3^2+560\right),\\
\label{DS7c}
v_\pm&\equiv&\frac{1}{\pm 4 \sqrt{(e-2)^2-3 s_3^2}-8 e-3 s_3^2+4}\;,\\
\label{DS7d}
u_2&\equiv& -\sqrt{3 v_+}\; K\left(8 \sqrt{(e-2)^2-3 s_3^2}\; v_+\right)\;,\\
\label{DS7e}
{\mathcal P}(z)&\equiv&{\mathcal P}(z; g_2,g_3)\;,\\
\label{DS7f}
P&=& 8 \sqrt{v_+}K\left( \frac{v_+}{v_-}\right)\;,\\
\label{DS7g}
z_{1,2}(t)&=&\frac{1}{2}\left( s_3\pm \sqrt{{\mathcal P}\left(\frac{i t\sqrt{3}}{2}+u_2\right)-{\mathcal P}(u_2)}\right)
\;.
\end{eqnarray}
$K$ denotes the complete elliptic integral of first kind,
see \cite{AS72} Ch. $17$.
The sign $\pm$ in (\ref{DS7g}) has to be chosen to fit with $z_{1,2}(t)$ according to the initial conditions. $z_{1,2}(t)$ performs
periodic oscillations about its mean value $\frac{s_3}{2}$ with period $P$ according to (\ref{DS7f}), see figure \ref{fig_z12}.\\

We now turn to the solution for $\phi_1(t)$. The conserved quantities (\ref{DS1}), (\ref{DS2}) can be used to express
$\dot{\phi_1}$ solely in terms of $z_1$:
\begin{equation}\label{DS8}
\dot{\phi_1}= \frac{(e-2) z_1+2 s_3}{z_1^2-1}
\;.
\end{equation}
Since $z_1(t)$ is a periodic function, $\phi_1(t)$ will also be periodic in time, except for a constant drift that
moves $\phi(t)$ with a certain amount $\delta \phi$ during one period $P$. This is illustrated in figure \ref{fig_phi1}.
Moreover, it turns out that $\frac{d\phi_1}{d z_1}$ can be written as a function of $z_1$ that is the quotient of a rational function
and a square root of a polynomial of $4$th degree. Hence  $\phi_1(z_1)$
is expressible in terms of elliptic integrals and, after inserting $z_1(t)$,
an explicit form of $\phi_1(t)$ is possible, analogously for $\phi_2(z_2)$.
We defer the details and the final result to Appendix \ref{sec:C}.

\subsection{Solutions close to the ground states}\label{sec:DL}

The configuration $({\mathbf s}_1,{\mathbf s}_2)$ with minimal energy (\ref{DS2a}) under the constraints
$|{\mathbf s}_1|^2=|{\mathbf s}_2|^2=1$ is a {\em critical point} of (\ref{DS2a}), i.~e.~, it satisfies the conditions
\begin{eqnarray}\label{DL1a}
-\nabla_{{\mathbf s}_1}H&=& 3 {\mathbf s}_2\cdot{\mathbf e}\;{\mathbf e}-{\mathbf s}_2=\lambda_1\;{\mathbf s}_1\;,\\
\label{DL1b}
-\nabla_{{\mathbf s}_2}H&=& 3 {\mathbf s}_1\cdot{\mathbf e}\;{\mathbf e}-{\mathbf s}_1=\lambda_2\;{\mathbf s}_2\;,
\end{eqnarray}
where $\lambda_1,\;\lambda_2$ are Lagrange parameters due to the constraints. Upon forming the scalar product
of both equations with ${\mathbf e}$ one easily derives the following alternative:
Either ${\mathbf s}_1\cdot{\mathbf e}={\mathbf s}_2\cdot{\mathbf e}=0$ or $\lambda_1\lambda_2=4$ and ${\mathbf s}_i=\pm{\mathbf e},\; i=1,2$.
In the first case, $H={\mathbf s}_1\cdot{\mathbf s}_2\ge -1$, whereas in the second case
the function $H$ assumes the ground state energy $h_0=-2$ if ${\mathbf s}_1={\mathbf s}_2 = \pm {\mathbf e}$.
Hence the two ferromagnetic configurations parallel to ${\mathbf e}$ constitute the ground states of the dipole pair.\\

The energy barrier between the two ground states has the value $\Delta E =1$. This can be seen as follows.
Any path $\pi$ in phase space joining the two ground states has at least one local energy maximum of height $h(\pi)$.
The minimum $h_1$ of $h(\pi)$ among all such paths $\pi$ is necessarily assumed at a saddle point and hence at a critical
point of (\ref{DS2a}). From the above classification of critical points only the possibilities
${\mathbf s}_1\cdot{\mathbf e}={\mathbf s}_2\cdot{\mathbf e}=0$ remain as candidates for saddle points and in this set
only the configurations with ${\mathbf s}_1=-{\mathbf s}_2$ assume the minimal energy $h_1=-1$. Hence $\Delta E = h_1-h_0 =1$.\\

For energies slightly above $h_0=-2$ it is sensible to linearize the eqm. Writing
\begin{eqnarray}\label{DL2a}
{\mathbf s}_1&=&
\left(
\begin{array}{l}
X_1\\
X_2\\
-1
\end{array}
\right)+{\mathcal O}(|{\mathbf X}|^2)
\;,\\
\label{DL2b}
{\mathbf s}_2&=&
\left(
\begin{array}{l}
X_3\\
X_4\\
-1
\end{array}
\right)+{\mathcal O}(|{\mathbf X}|^2)
\;,
\end{eqnarray}
we obtain the linearized eqm in the form
\begin{equation}\label{DL3}
\dot{\mathbf X}(t)=A\;{\mathbf X}(t)
\;,
\end{equation}
where ${\mathbf X}=(X_1,X_2,X_3,X_4)$. The matrix $A$ has the form
\begin{equation}\label{DL4}
A=\left(
\begin{array}{cccc}
 0 & 2 & 0 & 1 \\
 -2 & 0 & -1 & 0 \\
 0 & 1 & 0 & 2 \\
 -1 & 0 & -2 & 0
\end{array}
\right)
\;,
\end{equation}
and its eigenvalues are $\pm i, \pm 3 i$. For later purposes we write down the first two components of the solutions
of (\ref{DL3}) using the initial conditions $X_i(0)=x_i,\; i=1,\ldots,4$.
\begin{eqnarray}\nonumber
X_1(t)&=&\frac{1}{2} \left((x_1-x_3) \cos (t)+(x_2-x_4) \sin (t)+\right.\\
\label{DL5a}
&&\left.(x_1+x_3)\cos (3 t)+(x_2+x_4) \sin (3 t)\right)\;,\\
\nonumber
X_2(t)&=&\frac{1}{2} \left((x_2-x_4) \cos (t)+(x_3-x_1) \sin (t)+\right.\\
\label{DL5b}
&&\left.(x_2+x_4)\cos (3 t)-(x_1+x_3) \sin (3 t)\right)
\;.
\end{eqnarray}
From this we can calculate the lowest non-trivial order of
\begin{eqnarray}\label{DL6a}
z_1(t)&=&\pm\,\sqrt{1-(X_1(t)^2+X_2(t)^2)}\\
\label{DL6b}
&=&-1+\frac{1}{2}(X_1(t)^2+X_2(t)^2)+{\mathcal O}(|{\mathbf X}|^4)\\
\nonumber
&=&-1+\frac{1}{4} \left( (x_1^2+x_2^2+x_3^2+x_4^2)+\right.\\
\nonumber
&&
\left(x_1^2+x_2^2-x_3^2-x_4^2\right)\cos(2 t)+\\
\label{DL6c}
&& \left.2(x_1  x_4- x_2 x_3)\sin(2 t)
  \right)+{\mathcal O}(|{\mathbf X}|^4)
\;.
\end{eqnarray}
At first sight it is remarkable that $z_1(t)$ contains no term proportional to $\sin(6 t)$ or $\cos(6 t)$
as one would expect from the possible addition of frequencies in $X_1(t)^2+X_2(t)^2$. However, the result
(\ref{DL6c}) is in accordance with the low energy limit of the exact solution (\ref{DS7e}) of $z_1(t)$.
Hence in the low energy limit ${\mathbf s}_1(t)$ performs a harmonic oscillation with the two angular frequencies
$\omega_1=1$ and $\omega_2=3$ in the $x-y-$plane and $\omega_3=2$ in the $z-$direction.
Recall that according to (\ref{DE5b}) we have chosen the unit of angular frequency to be $\omega_0$.\\

\section{Thermodynamics}\label{sec:T}
A direct experimental test of the results of section \ref{sec:D} for
nanomagnets  is naturally affected by thermal fluctuations due to finite temperatures.
Hence it seems worth while to investigate the thermodynamics of
magnetic dipoles, especially to calculate thermodynamic functions such as the specific heat and the susceptibility.
Furthermore, we will consider the autocorrelation function ($ac$) in the low temperature limit.
The theoretical results will be compared with those of simulations of the system of two magnetic dipoles coupled to a
heat bath. The methods used are described in the following subsection.

\subsection{Methods}\label{sec:M}
As it is well-known, thermodynamic functions such as the specific heat and the susceptibility
can be derived from the partition function $Z(\beta)$ of the system. However, we were not able
to explicitly calculate $Z(\beta)$ for the Hamiltonian (\ref{DS2}). Fortunately, there exist
powerful approximation schemes to overcome this difficulty. On the one hand it is possible to
derive the moments of $H$ and thus the complete high temperature expansion (HTE)
series of $Z(\beta)$.
A large order truncation ($n=100$) together with an appropriate Padé approximation
then yields very accurate approximations of $Z(\beta)$
and hence of the specific heat $c(\beta)$ down to low temperatures. On the other hand, the integrals
over the $4$-dimensional phase space defining $Z(\beta)$ can be transformed conveniently to allow for a low temperature
asymptotic expansion (LTA) of several orders of, say, $n=12$.
The domains of validity of the two approximations, HTE and LTA, overlap, therefore
together provide an accurate
approximation of $c(\beta)$ without any need of interpolation.

Analogous remarks apply to the zero field susceptibility $\chi(\beta)$. Here it is possible to combine the complete
HTE series with an LTA of several orders.
Since the easy axis susceptibility $\chi(T)$ diverges for $T\rightarrow 0$
with the power $T^{-1}$ it is more appropriate to
plot the product $T\;\chi(T)$ as a function of $T$. In contrast to this, the hard axis susceptibility
approaches a finite value for $T\rightarrow 0$.
The investigation of the autocorrelation function $ac$ and its thermal average $\langle ac \rangle$
combines dynamical and thermodynamical aspects of the system under consideration. As mentioned above, we
will restrict ourselves to the low temperature asymptotic expansion up to terms of first order in $T$.
In this realm it is sufficient to consider the solutions
of the eqm close to the ground states, see subsection \ref{sec:DL}, and to perform the integrations
within the ``harmonic oscillator approximation", i.~e.~an approximation of the Hamiltonian that is quadratic in the
deviations from the ground state.

Furthermore, we have used classical spin dynamics and Monte Carlo simulations in order to compare our analytical derivations with numerical results.

\subsection{Partition function}\label{sec:Z}
\subsubsection{LTA}
As a first step we derive the low temperature asymptotic expansion (LTA) of the partition function $Z(\beta)$, where $\beta$
is the dimensionless inverse temperature
\begin{equation}\label{Z1}
\beta=\frac{E_0}{k_B T}
\;,
\end{equation}
and the energy unit $E_0$ has been defined in (\ref{DS2a}).
We will also use the dimensionless temperature $\frac{k_B\,T}{E_0}$ which again will be denoted by $T$
without danger of confusion.
According to its definition,
\begin{eqnarray}\nonumber
Z(\beta)&=&\frac{1}{(4\pi)^2}\int_{-1}^1dz_1\int_{-1}^1dz_2\int_0^{2\pi}d\phi_1\int_0^{2\pi}d\phi_2 e^{-\beta H}.\\
&&\label{Z2}
\end{eqnarray}
For fixed $\phi_2$ we substitute $\phi_1=\phi+\phi_2$ and obtain the partial integral
\begin{eqnarray}\nonumber
{\mathcal I}_1&\equiv&
\int_0^{2\pi}d\phi_1\int_0^{2\pi}d\phi_2\; e^{-\beta H}\\
\nonumber
&=& 2\pi e^{2\beta z_1 z_2}\int_0^{2\pi}d\phi \;e^{-\beta \sqrt{(1-z_1^2)(1-z_2^2)}\cos\phi}\\
\label{Z3}
&=& (2\pi)^2 e^{2\beta z_1 z_2} I_0\left(\beta \sqrt{(1-z_1^2)(1-z_2^2)}\right)\;,
\end{eqnarray}
where $I_n$ is the modified Bessel function of $n$th order.
Next we substitute $z_i=-1+u_i^2,\; u_i\ge 0,\;i=1,2,$ and obtain
\begin{eqnarray}\nonumber
Z(\beta)&=&
\int_0^{\sqrt{2}}u_1 du_1\int_0^{\sqrt{2}}u_2 du_2 \\
&&\nonumber
\exp\left(2\beta(-1+u_1^2) (-1+u_2^2)\right)\\
\label{Z4}
&& I_0\left(2\beta u_1 u_2 \sqrt{(1-\frac{1}{2}u_1^2)(1-\frac{1}{2}u_2^2)}  \right)
\;.
\end{eqnarray}
Now we consider the limit $\beta\rightarrow\infty$  by introducing polar coordinates
$u_1= \frac{r}{\sqrt{\beta}}\cos\psi,\;u_2= \frac{r}{\sqrt{\beta}}\sin\psi,$
extracting the factor $e^{2\beta}/\beta^2$ and evaluating the remaining integral only
in $0$th order of its Taylor series in $\beta^{-1}$. The domain of integration
is extended to the whole first quadrant. This gives the contribution to $Z$ in the
limit  $\beta\rightarrow\infty$ from the neighborhood of the ground state $z_1=z_2=-1$.
In order to include the equal contribution from the ground state $z_1=z_2=1$ we have to
insert a factor $2$.
We thus obtain the following asymptotic limit
\begin{eqnarray}\nonumber
Z(\beta)&\sim&
\frac{2\,e^{2\beta}}{\beta^2}\int_0^\infty r^3\,e^{-2r^2}\, dr\\
&&\label{Z5a}
\int_0^{\pi/2} d\psi\frac{1}{2}\sin(2\psi)
I_0(r^2\sin(2\psi))\\
\label{Z5b}
&=&
\frac{2\,e^{2\beta}}{\beta^2}\int_0^\infty r^3\,e^{-2r^2}\frac{\sinh(r^2)}{2r^2}\, dr\\
\label{Z5c}
&=&\frac{e^{2\beta}}{6 \beta^2}
\;.
\end{eqnarray}

The method can be extended to obtain the first terms of an asymptotic series expansion
for $Z(\beta)$. We omit the details and state the following result:
\begin{eqnarray}\nonumber
Z(\beta)&\sim&
e^{2 \beta } \left(\frac{1}{6 \beta ^2}+\frac{1}{9 \beta ^3}+\frac{1}{6 \beta ^4}+\frac{11}{27 \beta ^5}+\frac{227}{162
\beta ^6}\right).\\
&&\label{Z6}
\end{eqnarray}

\subsubsection{HTE}
Let us denote by $\mbox{Tr}(f)$ the integral of a function $f$ over $4$-dimensional phase space divided by
its volume $(4\pi)^2$. Then the HTE of $Z(\beta)$ reads
\begin{equation}\label{Z7}
Z(\beta) = \mbox{Tr} \left( e^{-\beta H}\right)=\sum_{n=0}^\infty \frac{(-\beta)^n}{n!}\,\mbox{Tr}(H^n)
\;,
\end{equation}
where $H$ is the Hamiltonian (\ref{DS2a}). With the aid of computer algebraic software we calculate the
moments $\mbox{Tr}(H^n)$ and hence the HTE of $Z(\beta)$ with the result
\begin{equation}\label{Z8}
Z(\beta) = \sum _{n=0}^{\infty } \frac{4^n \, F\left(1,-n;\frac{1}{2}-n;\frac{1}{4}\right) (-\beta )^{2 n}}{(2
   n+1)^2 (2 n)!}
\;,
\end{equation}
where $F(a,b;c;z)$ denotes the hypergeometric function, see \cite{AS72}, ch.~$15$.
Since $\lim_{n\rightarrow\infty}F\left(1,-n;\frac{1}{2}-n;\frac{1}{4}\right)=\frac{4}{3}$ the
radius of convergence of (\ref{Z8}) is the same as that of the exponential series, namely $r=\infty$.
Perhaps this explains the high quality of the approximations schemes based on (\ref{Z8}).

\subsection{Specific heat}\label{sec:CH}

\begin{figure}
\begin{center}
\includegraphics[clip=on,width=85mm,angle=0]{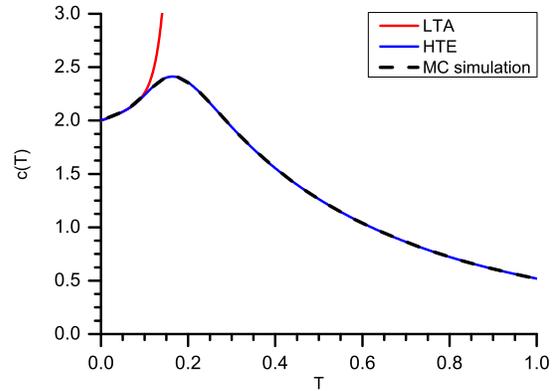}
\end{center}
\caption{Plot of the specific heat $c(T)$ vs.~$T$. The red curve shows the LTA
up to  order $T^{12}$; the blue curve is the result of using a $(50,50)$-Padé approximant based on the HTE of $c(\beta)$.
Both curves coincide for $0<T<0.08$ with a maximal relative deviation of $2\times 10^{-3}$.
The black dashed curve shows the results from our Monte Carlo simulation.
\label{fig_CT}
}
\end{figure}

According to the definition of the dimensionless specific heat
\begin{equation}\label{CH1}
c(\beta)=\beta^2\;\frac{\partial^2}{\partial \beta^2}\;\log Z(\beta)
\;,
\end{equation}
the approximations of $Z(\beta)$ based on HTE and LTA can be transferred to $c(\beta)$.
Especially, we apply a symmetric Padé approximation to the truncation of its
HTE of order $n=100$ . This coincides with a $12-$th order LTA of $c(T)$, the first five terms of which are
\begin{equation}\label{CH2}
c(T)=2 + \frac{4}{3}T +\frac{14}{3}T^2 + \frac{608}{27}T^3 + \frac{10810}{81}T^4+\ldots
\;,
\end{equation}
in the domain $0<T<0.08$ up to a relative deviation of $2\times 10^{-3}$ , see figure \ref{fig_CT}.
Alternatively,  the  specific heat can be obtained by numerically calculating
the fluctuations of the (dimensionless) total energy $E$ according to
\begin{equation}\label{CH3}
c^*(T)=\frac{1}{T^2} \left( \left< E^2 \right> - \left< E \right>^2 \right) \;
\end{equation}
by means of Monte Carlo simulations.

\subsection{Susceptibility}\label{sec:Chi}
\subsubsection{Easy axis}\label{sec:Chi_easy}

\begin{figure}
\begin{center}
\includegraphics[clip=on,width=85mm,angle=0]{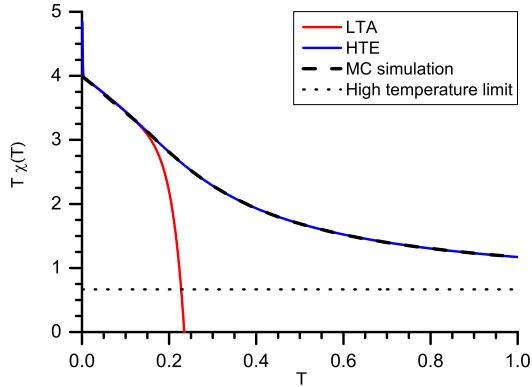}
\end{center}
\caption{Plot of the product $T\chi(T)$ vs.~$T$ for the easy axis. The red curve shows
the LTA up to  order $T^{12}$; the blue curve is a $(50,50)$-Padé
approximation based on the HTE of $\chi(\beta)$. Both curves coincide
for $0.01<T<0.06$ with a maximal relative deviation of $10^{-6}$.
The black dashed curve shows the results from our Monte Carlo
simulation. The dashed line represents the high temperature limit
$2/3$ of $T\chi(T)$. \label{fig_TChi} }
\end{figure}

The dimensionless zero field susceptibility
for infinitesimal magnetic fields in the direction ${\mathbf e}$ joining the two dipoles
(the ``easy axis")
is defined by
\begin{equation}\label{Chi1}
\chi(\beta)= \beta \langle S_3^2 \rangle =\beta \frac{\mbox{Tr}\left(S_3^2\,\exp(-\beta H)\right) }{Z(\beta)}
\;.
\end{equation}
The HTE of the numerator is
\begin{equation}\label{Chi2}
\mbox{Tr}\left(S_3^2\,\exp(-\beta H)\right)= \sum_{n=0}^\infty  \mbox{Tr}(S_3^2 H^n)\frac{(-\beta)^n}{n!}
\;.
\end{equation}
Again we can explicitly determine all moments occurring in (\ref{Chi2})
\begin{displaymath}
\mbox{Tr}(S_3^2 H^m)=
\end{displaymath}
\begin{equation}
\left\{
\begin{array}{ll}
\frac{2^{2 n+1} \, F\left(1,-n;-n-\frac{1}{2};\frac{1}{4}\right)}{4 n (n+2)+3}& \mbox{if } m=2n,\\
-\frac{2^{2 n+1} \left(n \, F\left(1,1-n;\frac{1}{2}-n;\frac{1}{4}\right)+4 n+2\right)}{(2 n+1) (2 n+3)^2}
& \mbox{if } m=2n+1.
\end{array}
\right.\\
\label{Chi3}
\end{equation}
and perform an HTE approximation of $T\chi(T)$ analogously to that of the specific heat.
The LTA of $T\chi(T)$ has been calculated up to $12-$th order, the first six terms being
\begin{equation}\label{Chi4}
T\chi(T)=4-\frac{16}{3}T-\frac{14}{9}T^2-\frac{140}{27}T^3-\frac{1628}{81}T^4- \frac{23888}{243}T^5-\ldots
\end{equation}
The combination of HTE and LTA results yields the form of $T\chi(T)$ displayed in figure \ref{fig_TChi}.
By means of Monte Carlo simulations, we obtain the dimensionless susceptibility
by evaluating the fluctuations of the total magnetization according to
\begin{equation}\label{Chi5}
\chi^*(T)=\frac{1}{T} \left( \left< {\mathbf M}^2 \right> - \left< {\mathbf M} \right>^2 \right).
\end{equation}

\begin{figure}
\begin{center}
\includegraphics[clip=on,width=85mm,angle=0]{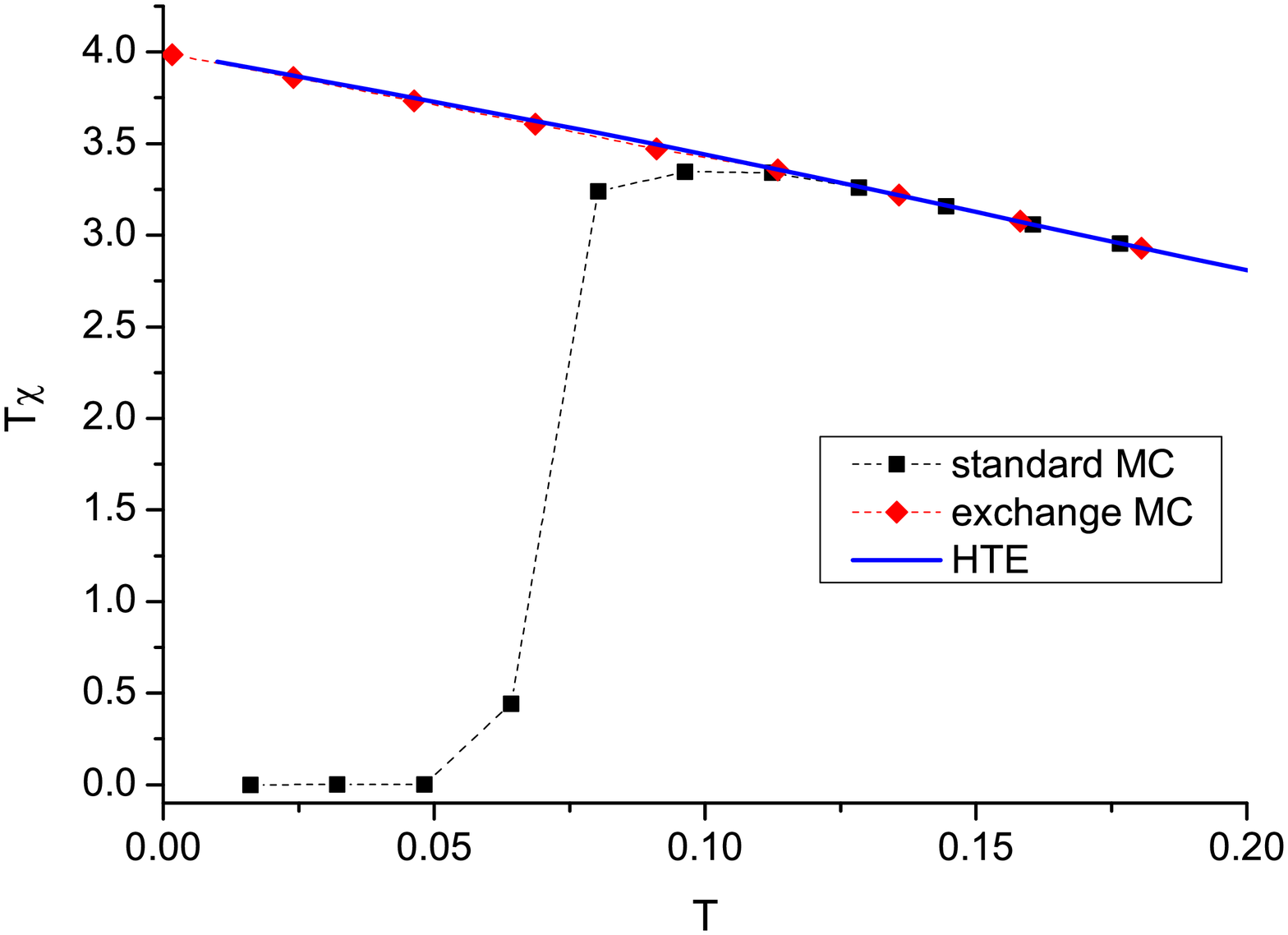}
\end{center}
\caption{Plot of the product $T\chi(T)$ vs.~$T$ for the easy axis. The blue curve shows
the Padé approximation based on the HTE of $\chi(\beta)$.
For $T<0.1$ the standard Monte Carlo simulation (black symbols) produces wrong results due to the non-ergodic behavior
of the bi-stable dipole system whereas the exchange Monte Carlo method (red symbols) reproduces the analytical results. \label{fig_TChi_ne} }
\end{figure}

It is interesting to note that in contrast to the specific heat the susceptibility at very low temperatures cannot be
determined correctly by using the standard Metropolis algorithm.
As a result of the dipolar interaction an inherent easy-axis anisotropy in the direction
of the connecting line between the two dipoles is formed resulting in a bi-stable system.
As pointed out in section~\ref{sec:DL}
at low temperatures the two dipoles are fluctuating
around their two possible ferromagnetic ground states that are separated by an energy barrier of $\Delta E=1$.
During the timescale of a typical computer simulation the two dipoles will be trapped in one of the directions;
any attempt to change both dipoles from one ground state configuration to the other is
rejected in most cases leading to non-ergodic behavior.
This is demonstrated in figure \ref{fig_TChi_ne}.
In contrast to the analytical results (blue curve) the numerically determined
susceptibility drops to zero for temperatures $T<0.1$.

This can be understood from the following argumentation: According
to equation (\ref{Chi5}) we expect $T \chi \to 4$ for $T \to 0$,
because of the $z$-component of the total magnetization $M_z = 1+1$
or $M_z=-1-1$ for each of the ground states ($M_x$ and $M_y$ are
both zero). This is the variance (fluctuation) of the total
magnetization $\textbf{M}$ since $\langle\textbf{M}\rangle = 0$ in
the ground state. The latter is only valid in a simulation if both
ground states are equally often generated such that the average of
$\textbf{M}$ becomes 0. If the system gets trapped in one of the
ground states we find $\langle\textbf{M}\rangle = \pm 2$ and hence
the variance vanishes according to $T \chi = 4-2\cdot 2=0$.

In order to obtain correct results we have used the so-called Exchange Monte Carlo method
\cite{H96} in which many replicas of the system with different temperatures are
simultaneously simulated and a virtual process exchanging configurations of these replicas is introduced.
This exchange process allows the system at low temperatures to escape from a local minimum,
hence leading to ergodic behavior and therefore producing correct data for the susceptibility
(shown as red symbols in figure \ref{fig_TChi_ne}).

Furthermore, it is interesting to note that the specific heat can be obtained
correctly by a standard Monte Carlo algorithm. In contrast to the susceptibility
which is calculated using the fluctuations of a \textit{directed} property,
e.~g.~the total magnetization, the specific heat is calculated by
sampling the fluctuations of the \textit{undirected} total energy.
Hence, the low temperature fluctuations in one of the two possible (degenerate)
ground states are sufficient to yield the correct statistics.

The same argumentation using the fluctations of a directed property
holds for the simulation of the hard axis susceptibility (see
subsection III D 2). However, for this direction there is no energy
barrier blocking the system.

\subsubsection{Hard axis}\label{sec:Chi_hard}

\begin{figure}
\begin{center}
\includegraphics[clip=on,width=85mm,angle=0]{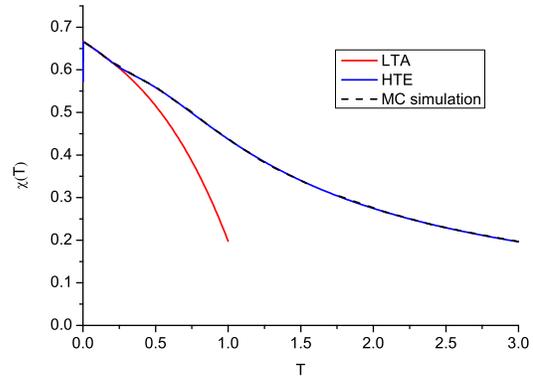}
\end{center}
\caption{Plot of the hard axis susceptibility $\chi(T)$ vs.~$T$. The blue curve shows
the Padé approximation based on the HTE of $\chi(\beta)$, the red curve the LTA according to
(\ref{Chi7}), and the black dashed curve the result of the Monte Carlo simulation.
\label{Fig13}
}
\end{figure}

The zero field susceptibility for the infinitesimal magnetic field in a direction perpendicular
to the line joining the two dipoles (the ``hard axis") will be calculated by the same methods as
for the easy axis. Without loss of generality we choose the $x$-axis as the hard axis.
Again we can explicitly determine all relevant moments
\begin{displaymath}
\mbox{Tr}(S_1^2 H^m)=
\end{displaymath}
\begin{equation}
\left\{
\begin{array}{l}
\frac{2^{2 n+1} \left(-(3 n+5) \, F\left(1,-n;-n-\frac{1}{2};\frac{1}{4}\right)+4 n+6\right)}{4 n (n+2)+3} \quad\mbox{ if } m=2n\,,\\
\frac{9\times 2^{2 n+1} \left((3 n+5) \, F\left(1,n+\frac{3}{2};n+1;4\right)+n+1\right) \Gamma \left(n+\frac{3}{2}\right)-i
   \sqrt{3 \pi } (3 n+5) n!}{18 (2 n+3) \Gamma \left(n+\frac{5}{2}\right)}\\
 \mbox{if } m=2n+1.
\end{array}
\right.\\
\label{Chi6}
\end{equation}
and obtain from this the HTE of the susceptibility and a corresponding $(50,50)$-Padé approximant that can
be used down to low temperatures of, say, $T=0.01$. The LTA leads to the terms
\begin{equation}\label{Chi7}
\chi(T)=\frac{2}{3}-\frac{2 }{9}T-\frac{2 }{27}T^2-\frac{14 }{81}T^3+{\mathcal O}(T^4)
\;.
\end{equation}
Both approximations can be combined and yield a result that is very close to that obtained by Monte Carlo
simulations, see figure \ref{Fig13}. It is physically plausible that a small magnetic field in $x$-direction
only leads to a small additional magnetization relative to that of the ground state. Hence the susceptibility
is expected to approach a finite value for $T\rightarrow 0$. This is confirmed by the above result for the LTA
(\ref{Chi7}). For the same reason the complications in the Monte Carlo simulations mentioned above, see section
\ref{sec:Chi_easy}, do not occur.

\subsection{Autocorrelation function}\label{sec:AC}

The autocorrelation function $ac$ or rather its thermal average $\langle ac \rangle$
provide typical characteristics of a system under the influence of thermal fluctuations.
In our case we consider $ac={\mathbf s}_1(0)\cdot {\mathbf s}_1(t)$ (the result for the second dipole
would be identical) and
will exactly evaluate $\langle ac \rangle$ in the limit $\beta\rightarrow\infty$.
From section \ref{sec:DL} we know already that only the three frequencies
$\omega_1=1,\;\omega_2=3$ and $\omega_3=2$ will occur in the Fourier spectrum of low
temperature oscillations. Since
\begin{eqnarray}\nonumber
ac&=&\sqrt{(1-z_1(0)^2)(1-z_1(t)^2)}\cos(\phi_1(0)-\phi_1(t))\\
\label{AC1}
&&+z_1(0)z_1(t)
\;,
\end{eqnarray}
we expect that the contribution $\langle z_1(0)z_1(t)\rangle$ will be suppressed by thermal averaging
over all phase shifts of the $z_1$-oscillations. On the other hand,
$\langle\sqrt{(1-z_1(0)^2)(1-z_1(t)^2)}\cos(\phi_1(0)-\phi_1(t))\rangle$ will probably not vanish since
the phase shifts of the $x-y$-oscillations have been already canceled in the argument of the
$\cos$-function. This conjecture has to be confirmed by the detailed calculations.\\

These calculations can be simplified by the following consideration. The
transformation ${\mathbf s}_i \mapsto -{\mathbf s}_i,\; i=1,2,$ introduces a minus sign
in the eqm (\ref{DE6a}) and hence can be considered as a kind of ``time reversal".
However, it leaves the $ac$ invariant and hence $\langle ac \rangle$ will be also invariant
under time reversal. Consequently, the terms of $ac$ proportional to $\sin(t), \sin(2t)$
and $\sin(3t)$ will vanish in the thermal average and need not be calculated.\\

The calculation of $\langle ac(t) \rangle$ is based on the approximation of the Hamiltonian $H$
to terms at most quadratic in the deviations from a ground state.
We do not give the details but the main steps are sketched
in Appendix \ref{sec:D}. The final result reads:
\begin{equation}\label{AC2}
\langle ac\rangle =1-\frac{4}{3 \beta }+ \frac{3 \cos (t)+\cos (3 t)}{3 \beta }+{\mathcal O}(\beta^{-2})
\;.
\end{equation}
This shows that indeed the frequency $\omega=2$ of the $z_1$-oscillation is suppressed by thermal averaging and can at most
occur as contributions of order ${\mathcal O}(\beta^{-2})$.\\

\begin{figure}
\begin{center}
\includegraphics[clip=on,width=85mm,angle=0]{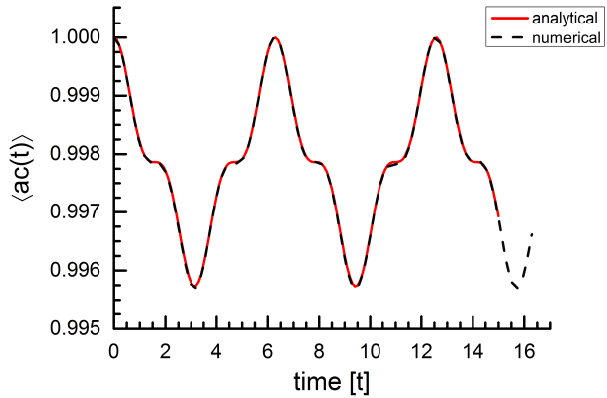}
\end{center}
\caption{Plot of the autocorrelation function $\langle ac(t)\rangle$
vs.~$t$ for a dimensionless temperature of $T=0.00160156$. The red
curve shows the analytical results; the black curve shows the
numerical results.
\label{fig_AC}
}
\end{figure}

\begin{figure}
\begin{center}
\includegraphics[clip=on,width=85mm,angle=0]{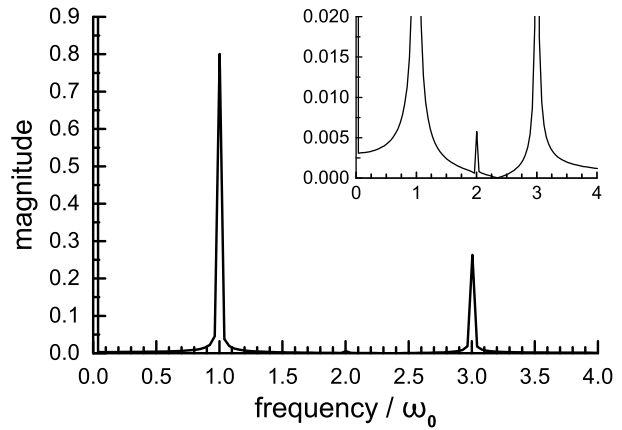}
\end{center}
\caption{Plot of the Fourier transform of the autocorrelation
function $\langle ac(t)\rangle$ vs.~$\omega$ for a dimensionless
temperature of $T=0.00160156$.
The inset shows the peak at $\omega =2$ which is almost suppressed by thermal averaging.
The amplitudes of the two large peaks are in the ratio of $2.95:1$
in accordance with (\ref{AC2}).
\label{fig_ACF}
}
\end{figure}

We compared these results with numerical simulations. In order to calculate the canonical ensemble
average numerically we used the so-called ``Gibbs approach" \cite{LL99}, where the trajectories
${\mathbf s}_1(t)$ for the dipoles are calculated for the \textit{isolated} system by solving
the equations of motion (\ref{DE6a}) and (\ref{DE6b}) over a certain number of time steps numerically.
The initial conditions for each trajectory are generated by a standard Monte Carlo simulation
for a temperature $T$. By averaging all generated trajectories at each time step one obtains the canonical ensemble average.
In figure \ref{fig_AC} we show a comparison of our analytical and simulation results in the time domain.
The Fourier transform of the simulation data (see figure \ref{fig_ACF}) yields the expected spectrum
showing three distinct peaks, where the peak at the frequency $\omega = 2$ is almost suppressed compared to the other peaks.

\section{Summary and Outlook}\label{sec:S}
In this paper we have investigated the system consisting of two magnetic dipoles, fixed in space
and interacting via its magnetic fields. The dynamics of this system has been completely resolved
and the general solution of the equations of motion has been given in terms of elliptic integrals
and Weierstrass elliptic functions. The thermodynamics of the two dipole system based on the canonical ensemble
has also been determined by means of series expansions, including the low temperature limit of the autocorrelation function.
The analytical results have been confirmed by numerical Monte Carlo simulations.

Hence we have found a simple but non-trivial example for a solvable system in the sense of classical mechanics
and of classical thermodynamics for systems with small particle numbers. The other motif of our studies was to prepare
the investigation of larger systems of interacting dipoles that have been recently realized by experimentalists.
Therefore it is in order to reflect about possible generalizations of our methods to larger systems.
First, it is clear that the Hamiltonian (\ref{DS5}) can be directly generalized to systems of $N$ dipoles and
yields the corresponding Hamiltonian eqm for the canonical coordinates $(p_i,\, q_i)=(\phi_i,\,z_i),\;i=1,\ldots,N$.
However, we do not expect that these eqm are completely integrable for $N>2$ due to the lack of a sufficient number of
integration constants. Nevertheless, it might be possible to find {\it some} exact solutions for larger systems of dipoles
and to identify its ground states. In particular, the linearization of the eqm close to the ground state(s) should be
possible and would only be practically limited by the size of $N$.
Concerning thermodynamics, we are pessimistic about the possibility to generalize our series expansions to larger systems
due to the complexity of the calculations. However, the ``linear oscillator approximation" would still be possible and
would yield low temperature limits of, e.~g.~, the autocorrelation function. In view of these difficulties the role of
numerical simulations would become more important for larger systems of magnetic dipoles.

\section*{Acknowledgment}
E.~H.~and C.~S.~acknowledge financial support from the equal
opportunity commissioner of the Bielefeld University of Applied
Sciences. H.-J.~S.~is indebted to Hans-Werner Sch\"urmann for
discussions about Weierstrass elliptic functions.

\appendix

\section{Elliptic integrals and Weierstrass elliptic functions}\label{sec:A}

There are many problems in theoretical physics that lead to elliptic integrals (EI)
or their inverses, elliptic functions (EF). We only mention a few:
\begin{itemize}
\item Various problems of classical mechanics \cite{B09} including one-dimensional motion
of a particle in a cubic or quartic potential, the spherical pendulum or the spinning top,
\item the magnetic field of a circular current loop \cite{J99}, Ch.~5,
\item the TE field in a slab filled with a Kerr non-linear medium \cite{S95},
\item certain solutions of the Korteweg-de-Vries equation \cite{B09}, and
\item problems from cosmology \cite{AW10}.
\end{itemize}
Nevertheless, most authors of physics textbooks seem to refrain from the use of these special functions,
one exception being the above-cited \cite{J99}. This is the more regrettable since by utilizing computer algebra software
both EI and EF can be evaluated with the same ease as, say, the $\sin$ and $\arcsin$ functions.\\

Here we cannot give an extended introduction into the field but will rather sketch the fundamental ideas.
One can understand the EI and EF as generalizations of the well-known ``circular case", where one
encounters the elementary integral
\begin{equation}\label{A1}
t=\int\; \frac{dx}{\sqrt{1-x^2}}= \arcsin x\;+\,t_0
\;,
\end{equation}
defined for $-1\le x \le 1$ and its inverse function
\begin{equation}\label{A2}
x(t)=\sin(t-t_0)
\;,
\end{equation}
that can be extended to a periodic function defined for all $-\infty < t < \infty$.
The following generalization of (\ref{A1}) is the {\em incomplete EI of the first kind}:
\begin{equation}\label{A3}
t=\int_0^x\; \frac{dx}{\sqrt{(1-x^2)(1-m x^2)}}\equiv F(\arcsin x,m)
\;.
\end{equation}
By the {\em complete EI of the first kind} one denotes the special case of the integral
\begin{equation}\label{A3a}
\int_0^1\; \frac{dx}{\sqrt{(1-x^2)(1-m x^2)}}\equiv K(m)
\;,
\end{equation}
that can be used, e.~g.~, for calculating the period of oscillation of a pendulum.\\

More generally, it can be shown \cite{A78}, Ch. 17, that any integral of a
rational function of $x$ and $\sqrt{P(x)}$, where $P(x)$ is a polynomial of at most
$4$th degree, can be expressed in terms of elementary functions and the so-called EI
of first, second or third kind. \\

Similarly as in the circular case, one is often interested in the function $x(t)$
rather than $t(x)$, that is, for the periodic extension of the inverse function of the EI, the EF.
There exist different versions of the EF; in this paper we will use the Weierstrass EF,
$u={\mathcal P}(z;g_2,g_3)$. It is first defined by inverting
\begin{equation}\label{A4}
z=\int_\infty^u\; \frac{dv}{\sqrt{P(v)}}
\;,
\end{equation}
where $P(v)=4 v^3-g_2 v-g_3$. Then ${\mathcal P}$ is extended to a doubly periodic
complex function, analytic in the whole complex plane except for the pole
at $z=0$ and its translates. For more details, see Chapter $18$ of \cite{A78} and
an introduction to the theory as it is given, e.~g.~, in \cite{B09} or \cite{B61}.

\section{Exact solution for $z_1(t)$}\label{sec:B}

The first step is to eliminate $z_2$ and $\phi_1 - \phi_2$ from (\ref{DS4c}) by using the
constants ${\mathbf S}\cdot{\mathbf e}=s_3$ and $H=e$. We write $z_1=z$. The result is

\begin{eqnarray}\label{B1a}
\dot{z}&=&\sqrt{-3 Q_+ Q_-}\;, \\
\nonumber \text{where}&&\\
\nonumber
Q_\pm &\equiv&\frac{1}{3} \left(1-2e -3 s_3 z + 3 z^2 \pm \sqrt{e^2-4 e-3 s_3^2+4}\right).\\
&&\label{B1b}
\end{eqnarray}
Upon substituting
\begin{eqnarray}\label{B2a}
v&=&(2 z-s_3)^2-v_0\;,\\
\label{B2b}
v_0&\equiv&\frac{2}{9} \left(8 e+3 s_3^2-4\right)
\end{eqnarray}
we obtain
\begin{equation}\label{B3}
t=\int \frac{dz}{\sqrt{-3 Q_+ Q_-}}=\frac{2}{\sqrt{-3}}\int \frac{dv}{\sqrt{4v^3-g_2 v-g_3}}\;,
\end{equation}
with $g_2$ and $g_3$ according to (\ref{DS7a}) and (\ref{DS7b}).
Inserting appropriate boundaries and writing $4v^3-g_2 v-g_3=P(v)$ we have
\begin{eqnarray}\label{B4a}
\frac{i \sqrt{3}}{2} t&=&\int_{-v_0}^{(2z-s_3)^2-v_0} \frac{dv}{\sqrt{P(v)}}\;,\\
\label{B4b}
&=&\int_{\infty}^{(2z-s_3)^2-v_0} \frac{dv}{\sqrt{P(v)}}-\int_{\infty}^{-v_0} \frac{dv}{\sqrt{P(v)}}\\
\label{B4c}
&\equiv& u_1-u_2
\;.
\end{eqnarray}
According to the definition of the Weierstrass ${\mathcal P}$-function, this is equivalent to
\begin{eqnarray}\label{B5a}
{\mathcal P}(u_1)&=&(2z-s_3)^2-v_0\;,\\
\label{B5b}
&=&(2z-s_3)^2+{\mathcal P}(u_2)
\;,
\end{eqnarray}
or, solving for $z$,
\begin{eqnarray}\label{B6a}
z(t)&=&\frac{1}{2}\left( s_3\pm \sqrt{{\mathcal P}(u_1)-{\mathcal P}(u_2)}\right)\;,\\
\label{B6b}
&=&\frac{1}{2}\left( s_3\pm \sqrt{{\mathcal P}\left(\frac{i \sqrt{3}}{2} t+u_2\right)-{\mathcal P}(u_2)}\right)
\;.
\end{eqnarray}
This confirms (\ref{DS7g}). As a consequence of choosing the lower boundary of the integral (\ref{B4a}) to be $-v_0$ we have $z(0)=s_3/2$.
For a more general solution one can simply replace $t$ in the r.~h.~s.~of (\ref{B6b}) by $t-t_0$.

\section{Exact solution for $\phi_{1,2}(t)$}\label{sec:C}

We write $z=z_{1,2},\; \phi=\phi_{1,2}$ and have to solve the integral
\begin{equation}\label{C0}
\int d\phi = \int \frac{\dot{\phi}}{\dot{z}} dz
\;,
\end{equation}
where $\dot{\phi}$ and $\dot{z}$ have to be inserted from
(\ref{DS8}) and (\ref{B1a}).\\
Defining
\begin{eqnarray}\label{C1a}
a_0&=&-s_3 \sqrt{-3 v_-}\;,\\
\label{C1b}
a_1&=&2 \sqrt{-3 v_-}\;,\\
\label{C1c}
\mu&=&\frac{1}{16} \left(4 e+s_3^2+4\right) \left(4 e+3 s_3^2-4\right)\;,\\
\label{C1d}
m&=&\frac{v_+}{v_-}
\;,
\end{eqnarray}
the substitution $x=a_1\,z +a_0$  yields
\begin{equation}\label{C2}
\frac{dz}{\dot{z}}=\frac{dz}{\sqrt{-3 Q_+ Q_-}}=\frac{dx}{a_1  \sqrt{\mu } \sqrt{\left(1-x^2\right) \left(1-m x^2\right)}}
\;.
\end{equation}
Upon this substitution (\ref{DS8}) can be written as
\begin{equation}\label{C3}
\dot{\phi}=\frac{(e-2) z+2 s_3}{z^2-1}=\frac{A_+}{1-n_+ x}+\frac{A_-}{1-n_- x}
\;,
\end{equation}
where
\begin{eqnarray}\label{C4a}
A_\pm&=&\frac{-2+e\mp 2s_3}{\pm 2+s_3}\;,\\
\label{C4b}
n_\pm&=&\mp \frac{1}{\sqrt{-3 v_-}(2\pm s_3)}\;.
\end{eqnarray}
These transformations lead to writing (\ref{C0}) as a sum of two integrals of the form
\begin{equation}\label{C5}
W\left(n;\left.x\right|m\right)\equiv
\int \frac{dx}{(1-n x)\sqrt{\left(1-x^2\right) \left(1-m x^2\right)}}
\;.
\end{equation}
Writing
\begin{equation}\label{C6}
\frac{1}{(1-n x)}=\frac{1}{1-n^2 x^2}+\frac{n\, x}{1-n^2 x^2}
\;,
\end{equation}
we obtain
\begin{eqnarray}\nonumber
&&W\left(n;\left.x\right|m\right)=\\
\nonumber
&&\Pi \left(n^2;\left.\sin ^{-1}(x)\right|m\right)+\\
\nonumber
&&\frac{n}{\sqrt{n^2-1} \sqrt{m-n^2}}
\tan ^{-1}\left(\frac{\sqrt{z^2-1} \sqrt{m-n^2}}{ \sqrt{m z^2-1}\sqrt{n^2-1}}\right)\;,\\
&&\label{B7}
\end{eqnarray}
where $\Pi$ is the incomplete elliptic integral of third kind, see
\cite{AS72} Ch.17. The final result hence reads
\begin{displaymath}
\phi(t)=\phi_0 + \frac{1}{a_1\sqrt{\mu}} \times
\end{displaymath}
\begin{equation}\label{C8}
\left( A_+ W\left(n_+;\left.\frac{z(t)-a_0}{a_1}\right|m\right) + A_- W\left(n_-;\left.\frac{z(t)-a_0}{a_1}\right|m\right)\right).
\end{equation}

\section{Low temperature limit of $\langle ac(t)\rangle$}\label{sec:D}

For the calculation of the low temperature limit of $\langle ac(t)\rangle$ we write
for the magnetic moments close to one of the ground states,
analogously to (\ref{DL2a}) and (\ref{DL2b}),
\begin{eqnarray}\label{D1a}
{\mathbf s}_1&=&
\left(
\begin{array}{l}
X_1\\
X_2\\
-1+\frac{1}{2}(X_1^2+X_2^2)
\end{array}
\right)
\;,\\
\label{D1b}
{\mathbf s}_2&=&
\left(
\begin{array}{l}
X_3\\
X_4\\
-1+ \frac{1}{2}(X_3^2+X_4^2)
\end{array}
\right)
\;,
\end{eqnarray}
and evaluate $H$ up to second order in $|{\mathbf X}|$. The result can be written as
\begin{equation}\label{D2}
H_2\equiv -2+{\mathbf X}\cdot {\mathbf M}\cdot{\mathbf X}
\;,
\end{equation}
where
\begin{equation}\label{D3}
{\mathbf M}=
\left(
\begin{array}{cccc}
 1 & 0 & \frac{1}{2} & 0 \\
 0 & 1 & 0 & \frac{1}{2} \\
 \frac{1}{2} & 0 & 1 & 0 \\
 0 & \frac{1}{2} & 0 & 1
\end{array}
\right)
\;.
\end{equation}
The eigenvalues of the symmetric matrix ${\mathbf M}$ are
$M_{1,2}=\frac{3}{2},\;M_{3,4}=\frac{1}{2}$. They are positive in accordance with
the fact that the considered ground state realizes the energy minimum $h_0=-2$. Their values are exactly
$1/2$ of the two basic frequencies $\omega_1=1,\;\omega_2=3$, i.~e.~,
of the absolute values of the eigenvalues of $A$, see (\ref{DL4}).
We perform a rotation into the eigenbasis of ${\mathbf M}$ and call the new coordinates
$Y_i,\,i=1,\ldots,4$.
In the second order approximation w.~r.~t.~$|{\mathbf X}|$ we then obtain the partition function
\begin{equation}\label{D4}
\frac{1}{2}Z(\beta) \sim e^{2\beta} \frac{1}{(4\pi)^2}\prod_{i=1}^4 \int_{-\infty}^\infty
\exp\left(-\beta M_i Y_i^2\right)\,dY_i
= \frac{e^{2\beta}}{12 \beta^2}
\;,
\end{equation}
which confirms the result (\ref{Z5c}) obtained by a different method. Recall that the factor
$\frac{1}{2}$ is introduced since the second ground state gives the same contribution to $Z(\beta)$.\\

The present method is also suited to calculate the low temperature limit of $\langle ac(t)\rangle$.
Consider first
\begin{eqnarray}\nonumber
ac_1(t)&\equiv& X_1(0)\,X_1(t)=\frac{1}{2} (Y_2-Y_4)\times\\
\nonumber
&& (-Y_1 \sin 3 t +Y_3 \sin t + Y_2 \cos 3 t-Y_4 \cos t).\\
&&\label{D5a}
\end{eqnarray}
If this expression is inserted into the integrals (\ref{D4}) only those terms
survive that are quadratic in the $Y_i$, namely
$\frac{1}{2} \left(Y_2^2 \cos 3 t+Y_4^2 \cos t\right)$.
Upon division by $\frac{1}{2}Z(\beta)$ we obtain
\begin{equation}\label{D6}
\langle ac_1(t)\rangle = \frac{2}{3\beta}\cos^3 t + {\mathcal O}(\beta^{-2})
\;.
\end{equation}
By azimuthal symmetry $\langle ac_2(t)\rangle\equiv \langle X_2(0)\,X_2(t)\rangle=\langle ac_1(t)\rangle$.
For $\langle ac_3(t)\rangle$ we have
\begin{eqnarray}\nonumber
\langle ac_3(t)\rangle&\equiv& \langle \quad(-1+\frac{1}{2}(X_1(0)^2+X_2(0)^2))\times\\
\label{D7a}
&&(-1+\frac{1}{2}(X_1(t)^2+X_2(t)^2))\quad\rangle\\
\nonumber
&=& 1 - \frac{1}{2}\langle (X_1(0)^2+X_2(0)^2)\rangle\\
\label{D7b}
&& -\frac{1}{2}\langle (X_1(t)^2+X_2(t)^2)\rangle + {\mathcal O}(\beta^{-2})
\end{eqnarray}
and by the same method as above it follows that the thermal average of the
time-dependent terms vanishes such that
\begin{equation}\label{D8}
\langle ac_3(t)\rangle =- \frac{4}{3\beta} + {\mathcal O}(\beta^{-2})
\;.
\end{equation}
Adding all contributions to $\langle ac(t)\rangle$ we obtain the following expression which proves (\ref{AC2}):
\begin{eqnarray}\label{D9a}
\langle ac(t)\rangle &=& 1-\frac{4}{3\beta} + \frac{4}{3\beta}\cos^3 t+{\mathcal O}(\beta^{-2})\\
\nonumber
&=&1-\frac{4}{3 \beta }+ \frac{3 \cos (t)+\cos (3 t)}{3 \beta }+{\mathcal O}(\beta^{-2}).\\
\label{D9b}
\end{eqnarray}


\end{document}